\documentclass{ws-procs975x65}
\def\b{\alpha}
\def\p{\phi}
\def\th{\theta}
\def\z{\zeta}

\def\htip{h_{{\rm tip}}}

\def\T{T_{\sigma s}}
\def\eend{\epsilon_e}
\def\es{\epsilon_*}
\def\cse{c_{se}}
\def\css{c_{s*}}
\def\Floc{f_{NL}^{loc}}
\def\Feq{f_{NL}^{eq}}
\def\rb{\beta}

\def\be{\begin{equation}}
\def\ee{\end{equation}}
\def\bea{\begin{eqnarray}}
\def\eea{\end{eqnnarray}}

\begin{document}

\title{COMBINED LOCAL AND EQUILATERAL NON-GAUSSIANITIES FROM MULTIFIELD DIRAC-BORN-INFELD INFLATION}

\author{SEBASTIEN RENAUX-PETEL}

\address{APC (UMR 7164, CNRS, Universit\'e Paris 7)\\
Paris, 75205 Paris Cedex 13, France\\
E-mail: renaux@apc.univ-paris7.fr}

\begin{abstract}
We explain the motivation and main idea of our work in Ref.~\refcite{RenauxPetel:2009sj}. We present a simple model of
 multifield Dirac-Born-Infeld inflation whose bispectrum exhibits a linear combination of the equilateral and local shapes, which are usually considered as separate possibilities. We also point out the presence of a particularly interesting component of the primordial trispectrum.
\end{abstract}

\keywords{Inflation; String theory; Non-Gaussianity.}

\bodymatter

\section{Introduction}

The study of the non-Gaussian properties of the primordial density fluctuations is a very promising tool with which to further discriminate between competing scenarios of the very early universe \cite{Komatsu:2009kd}. The departure from Gaussianity is usually measured by means of connected $n$-point correlation functions of the primordial curvature perturbation $\zeta$. Chief amongst these is the three point-function, or in Fourier space the bispectrum. The sum of the corresponding three momenta vanishes by virtue of translational invariance, thus forcing the momenta to form a triangle. The shape of the triangles for which the bispectrum signal is largest turns out to be a key discriminant between models \cite{Babich:2004gb}. In particular, in the two well known types of scenarios that can generate large non-Gaussianities -- multiple field models and single-field models with non-standard kinetic terms --  the bispectrum peaks respectively for squeezed (the so-called local shape) and equilateral triangles, their amplitude being respectively characterized by the parameters $\Floc$ and $\Feq$. The best motivated model that produces equilateral non-Gaussianities is the so-called Dirac-Born-Infeld (DBI) scenario \cite{Alishahiha:2004eh}. This is a string inspired model in which an extended object of three spatial dimensions (a D3-brane) evolves in the six extra dimensions of string theory. Its motion is governed by a non-standard Dirac-Born-Infeld action -- hence its name -- resulting in potentially large non-Gaussianities. Actually, the simplest models, in which the brane moves along a single radial direction, are already under strain because the predicted $\Feq$ is in excess of the observational bound \cite{Baumann:2006cd}. However, the brane can a priori move and fluctuate in each of the six extra dimensions, in which case a multifield description is required. It was shown in a model-independent way that the amplitude of equilateral non-Gaussianities in this more general framework is reduced compared to the single-field case \cite{Langlois:2008wt,Langlois:2008qf,Langlois:2009ej}, which is therefore of crucial importance for model-building. Here we report on an example of a multifield DBI inflationary scenario where it is possible to assess the importance of this supression. We also point out that, because of multiple fields effects, the large scale evolution can be highly nonlinear, resulting in possibly large local non-Gaussianities, thus giving the first example in which both equilateral and local non-Gaussianities -- usually considered as separate possibilities -- are present at an observable level in the same model. Their combined presence also manifests itself non-trivially in the connected 4-point function of $\zeta$, the trispectrum.

\section{A model of multifield DBI inflation and non-Gaussianities}

 In our scenario, inflation is still driven by a single inflaton scalar field $\phi$, namely the D3-brane solely moves along the standard radial direction of a throat. When the mobile D3-brane and an anti D3-brane sitting at the tip of the throat come within a string length, an open string mode stretched between them becomes tachyonic, triggering their annihilation and the end of inflation. As the brane-antibrane distance is \textit{six-dimensional}, it acquires some dependence upon the fluctuations of the light fields parametrizing the angular position of the brane. Hence the value of the inflaton at which the instability signals is modulated and the duration of inflation varies from one super-Hubble region to another (see Fig.~\ref{fig1}). In this way the angular, initially entropic, perturbations are converted into the curvature perturbation \cite{Lyth:2006nx}, its final value being different from the one at (sound) horizon crossing, denoted by a $*$:
\be
{\cal P_{\z}} = {\cal P_{\z*}}  \left( 1+\T^2 \right) \,.
\label{curvature}
\ee
\def\figsubcap#1{\par\noindent\centering\footnotesize(#1)}
\begin{figure}[b]%
\begin{center}
 \parbox{2.1in}{ \includegraphics[width=0.6\textwidth]{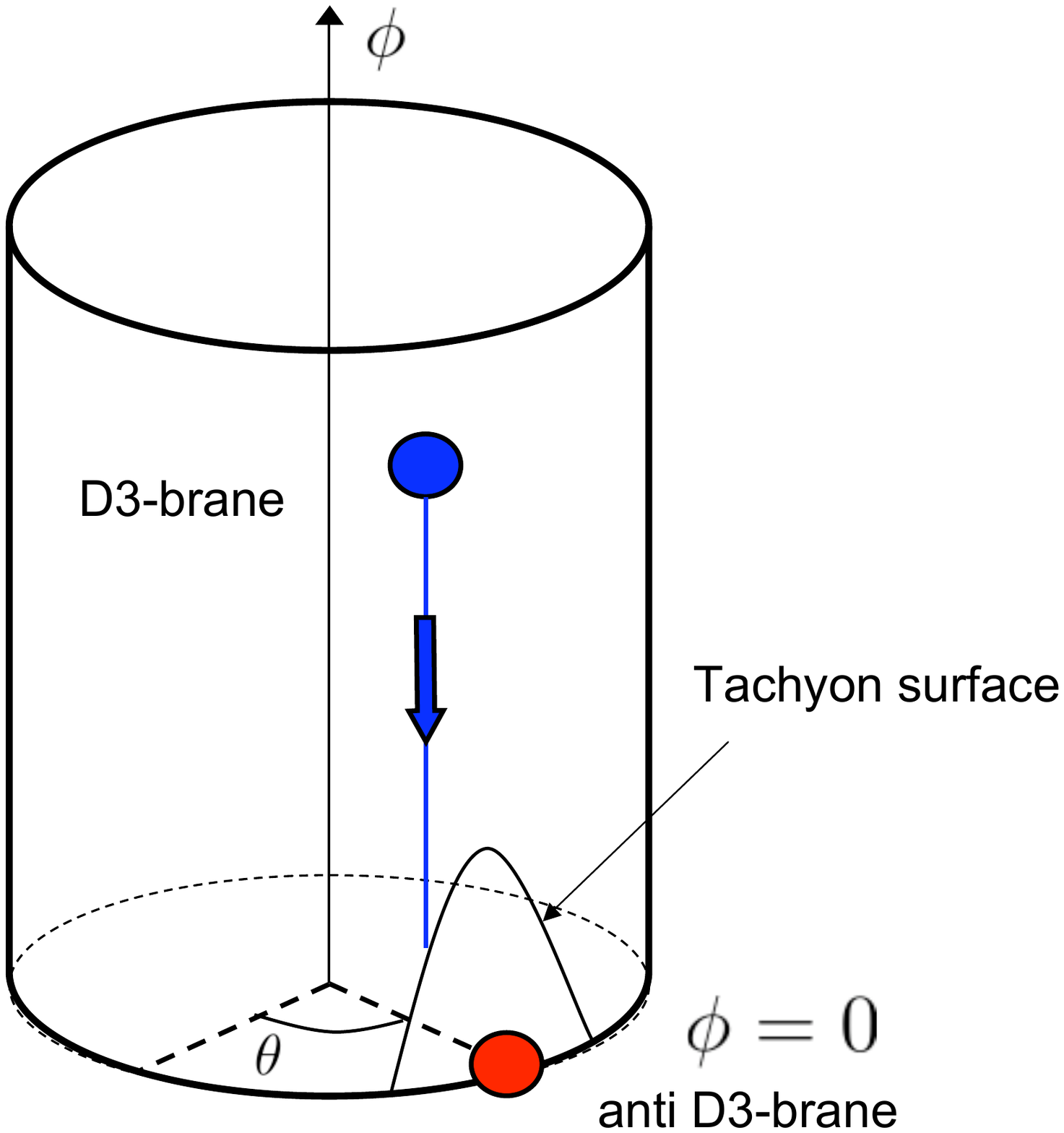}\figsubcap{a}}
 \hspace*{4pt}
 \parbox{2.1in}{\includegraphics[width=0.6\textwidth]{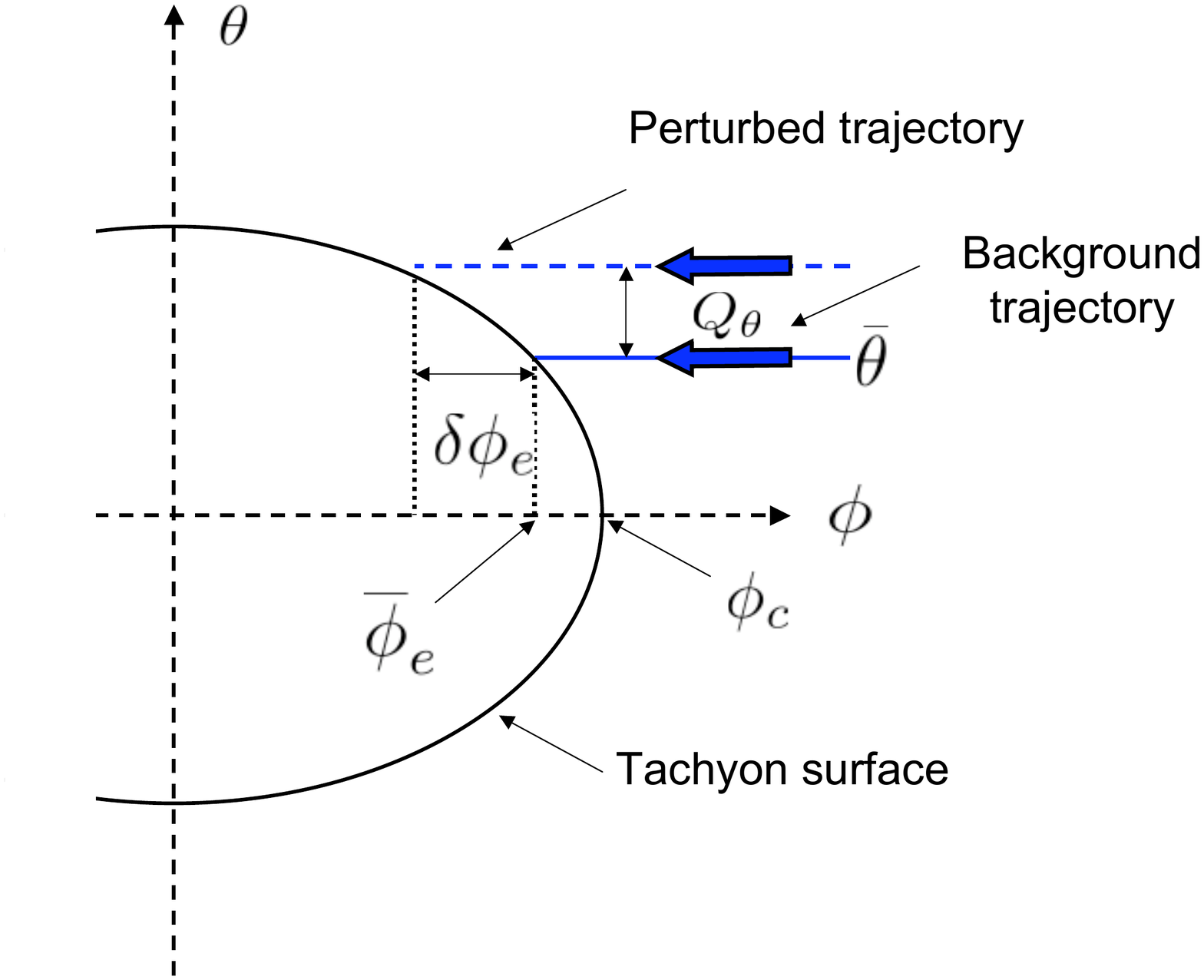}\figsubcap{b}}
 \caption{ (a) A simplified picture of the geometry at the tip of the throat, with one angular direction $\th$ only. The radial inflationary trajectory is represented by the blue line. (b) The tachyon surface, at which inflation ends, in the $\phi-\th$ plane. The end value of the inflaton is shifted from ${\overline \p_e}$ to ${\overline \p_e} +\delta \p_e$ due to the angular fluctuation $Q_{\th}$, hence the duration of inflation varies from one super-Hubble region to another.}
\label{fig1}
\end{center}
\end{figure}
The quantity $\T^2$ quantifies the efficience of the transfer. While it vanishes in single-field DBI inflation, in our scenario it is given by
\be
\T^2=\frac{\es}{\eend \cse\css} \tan^2(\b)\, \rb^2 
\label{transfer}
\ee
where the subscript $e$ denotes evaluation at the end of inflation and $\alpha$ and $0 < \rb \leq 1$ are geometrical factors associated to the position of the branes and the geometry of the throat. In the relativistic regime where $c_s \ll 1$, this transfer function can be very large, in which case the curvature perturbation is mostly of entropic origin.

At the level of the bispectrum, the entropic transfer diminishes the amount of equilateral non-Gaussianities with respect to the single field case \cite{Langlois:2008wt}:
\be
\Feq =-\frac{35}{108}\frac{1}{\css^2}\frac{1}{1+\T^2 }\,.
\label{f_NL3}
\ee
Concerning local non-Gaussianities, they can be large given that the end value of the inflaton is strongly nonlinearly related to the angular separation of the branes. In case where the end of inflation takes place in the relativistic regime $c_{s e} \ll 1$ and the entropic transfer is large $\T^2 \gg 1$, one indeed finds
   \be
f_{NL}^{loc}=\frac{5}{6}\frac{1}{\sin^2(\b) \cos(\b)} \frac{\htip m_s}{H_e}\,,
\label{FNL4-simplified}
\ee
where $m_s$ is the string mass and $h_{tip}$ is the warp factor at the tip of the throat. To avoid stringy corrections, one requires that at least $\frac{\htip m_s}{H_e} \gtrsim 1$ \cite{Frey:2005jk}, which therefore makes local non-Gaussianities observably large.
 
\section{Conclusion}

The common presence of non-standard kinetic terms and multiple-field effects implies that multifield DBI inflation is able to produce both large equilateral and local non-Gaussianities in the bispectrum. Moreover, due to the presence of light scalar fields with non standard kinetic terms, other than the inflaton, it can be shown that the trispectrum acquires a component with a particular momentum-dependence whose amplitude is given by the product $\Floc \, \Feq$ \cite{RenauxPetel:2009sj}. This consistency relation is valid for every DBI model, not restricting to a radial trajectory nor to a specific scenario for the entropy to curvature transfer. It thus constitutes an interesting observational signature of multifield DBI inflation which one can hope to test with forthcoming experiments.

\end{document}